# Optical properties of the $Bi^+$ center in $KAlCl_4$

A.A. Veber[1*], A.N. Romanov[2,3], O.V. Usovich[4], Z.T. Fattakhova[5], E.V. Haula[5], V.N. Korchak[5],
L.A. Trusov[4], P.E. Kazin[4], V.B. Sulimov[2,3] and V.B. Tsvetkov[1]

[1]*A.M. Prokhorov General Physics Institute, Russia Academy of Sciences, 38 Vavilov Str., 119991, Moscow, Russia*
[2]*Research Computer Center of M.V. Lomonosov Moscow State University, 1 Leninskie Gory, Build. 4, 119992 Moscow, Russia*
[3]*Dimonta Ltd.,15 Nagornaya Str., Build. 8, 117186 Moscow, Russia*
[4]*Department of chemistry, M.V.Lomonosov Moscow State University, 1 Leninskie Gory, Build.3 119991 Moscow, Russia*
[5]*N.N. Semenov Institute of Chemical Physics, Russian Academy of Sciences, 4 Kosygina Str., 119991, Moscow, Russia*

*Corresponding author. **E-mail**: alexveb@gmail.com, **Tel.**: +7(499)503-8274, **Fax**: +7(499)135-0270

*Abstract*
Temperature behavior of the $Bi^+$ center emission, excitation bands and luminescence decay law in $KAlCl_4$ crystal are investigated. Abrupt changes of the monocation optical properties are observed at phase transitions of the host. The observed optical transitions are assigned to the specific energy states of $Bi^+$ ion. It is shown that two thermalized levels are responsible for the luminescence. The experimental temperature behavior of the emission and excitation bands are in agreement with theory of electron-phonon interaction of impurity centers in solids; the effective-phonon frequencies and Huang-Rhys parameters are estimated. The configuration coordinate diagram is determined for the $Bi^+$ center in $KAlCl_4$ crystal.

*1. Introduction*
Near infrared (NIR) luminescence of Bi-doped media was discovered in 2001 by Fujimoto et al. in silicate glasses [1], and optical amplification was demonstrated later in this material [2]. To date lasing and optical amplification have been demonstrated in ~1.1-1.6 μm range using Bi-doped media [3-6]. Luminescence properties of various Bi-doped media: glasses [1, 7-10], crystals [11-16] as well as Bi-containing crystals [17-21] were investigated. Nevertheless the nature of Bi-related NIR luminescence is still subject of intense debates. A lot of various Bi-centers that could be responsible for the observed NIR luminescence were suggested: BiO, $Bi^+$, $Bi_2^{4+}$, $Bi_5^{3+}$, $Bi_8^{2+}$, $Bi_2^{2-}$, $Bi_2^-$, $Bi_4$, Bi-related defects etc. Often several luminescent species are observed simultaneously in a Bi-doped media [8, 10, 22-25]. To date only a few Bi-luminescent centers (e.g. $Bi^+$ [11, 26], $Bi_5^{3+}$ [17-20] and $Bi_8^{2+}$ [19]) were obtained solely.

Recently it was shown that Bi-doped $KAlCl_4$ possess intense NIR-luminescence near 1 μm [27]. By analysis of spectroscopic data this band was assigned to $Bi^+$ ion that isomorphically substitutes for $K^+$ in this crystal. Probably the luminescence of $Bi^+$ was observed in other crystalline hosts also [11, 12, 14, 16]. The Bi-doped $RbPb_2Cl_5$ [11] and $KMgCl_3$ [26] are the most prominent and reliable examples in this case. Frequently the NIR luminescence observed in Bi-doped glass media is ascribed to $Bi^+$ ion also [24, 28, 29]. Quite reliably it was demonstrated for $ZnCl_2/AlCl_3/BiCl_3$ glass [24].

In most cases the observed optical properties and NIR luminescence of Bi$^+$ ascribed to intraconfigurational transitions of an open external 6p$^2$ shell of the monocation, implying strong dependence of its optical properties on a host material. Nevertheless up to now there is no any detailed investigation on the Bi$^+$ optical center features and, in particular, its vibronic properties.

The results of further investigation of the Bi$^+$ center optical properties in KAlCl$_4$ crystal are reported in this paper. The temperature dependences of shapes, widths, positions, and intensities of emission and excitation bands as well as luminescence decay law over the temperature range of 77-410 K are investigated. The obtained experimental data are considered in terms of electron-phonon interaction of Bi$^+$ center with the host crystal.

## *2. Experimental*

The same as before [27] Bi-doped KAlCl$_4$ polycrystalline sample is used in present investigation. The details of the sample preparation as well as experimental setup description could be found in our previous works [24,27]. In brief, the Bi-doped KAlCl$_4$ specimens were synthesized by melt crystallization. To be sure that bismuth monocation is formed in melt, AlCl$_3$ was taken in excess and the controlled reduction of BiCl$_3$ by zinc metal was performed. The experimental setup has not been changed, only the temperature range is extended up to 410 K and additional measurements are carried out.

## *3. Results*
### *3.1. Luminescence. Experimental results.*

The luminescence spectrum of Bi$^+$ in KAlCl$_4$ crystal strongly depends on the temperature (fig.1). However the luminescence peak shape isn't only determined by usual changes like thermal broadening, peak shift etc. with changing of temperature only, but also by the phase transitions of the host. . It is known, that KAlCl4 exists in three polymorphic modifications: I – low temperature phase, II – room temperature and III – high temperature phase, stable up to the melting [30].

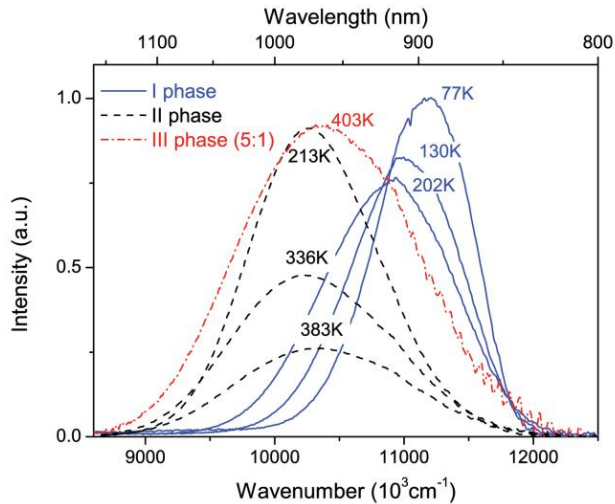

Fig.1. Luminescence spectra of Bi$^+$ in KAlCl$_4$ crystal at different temperatures that correspond to I, II and III phases of the host. Spectrum at 403 K is enlarged fivefold for clarity. Hereinafter a correct intensity function corresponds to wavenumber scale; the wavelength scale added for better perception.

Previously it has been shown, that abrupt changes in the luminescence decay time are observed during I↔II phase transition [27]. Further investigation revealed that the leap of the lifetime occurs at temperature of about 390 K during the II↔III phase transition (fig. 2a).

Leaps in temperature dependences of FWHM and peak position are observed during I↔II phase transition (fig. 2b). Somewhat unusual gradual (not sudden) changes are observed during the II↔III phase transition for these measurands.

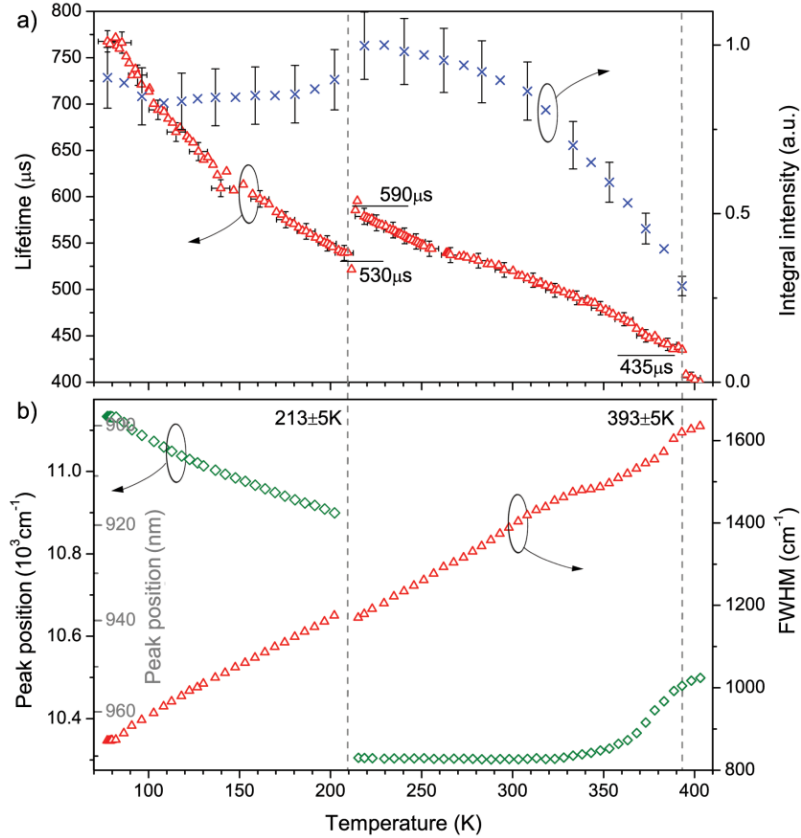

Fig.2. Behavior of the main features of $Bi^+$ luminescence over temperature: a) lifetime and integral intensity; b) peak position and FWHM. Phase transitions are marked with vertical dashed lines.

The obtained dependence of the integral luminescence intensity versus temperature shows that its value depends weakly on the temperature up to ~320 K. Moreover, within the experimental precision, it is a constant in whole temperature range of the phase I existence, while luminescence lifetime decreases with heating in this case. These facts indicate that the reason of the decay time decrease is a thermalization of some upper energy level and the probability of radiation transition from this level to the ground state is higher then for the bottom one.

### *3.2. Lifetime and integral intensity analysis.*

The present situation isn't unique and this effect was experimentally observed previously, e.g. for $Bi^{3+}$ [31] and $Cr^{3+}$ [32, 33] optical centers. In this case, ignoring degeneracy difference between two levels, the luminescence lifetime may be given by:

$$\tau_{is}(T) = \frac{1 + e^{-\Delta E/kT}}{\tau_s^{-1} + \tau_i^{-1} \cdot e^{-\Delta E/kT}} \quad (1),$$

$\tau_s$ - lifetime of lower ("storage") level;
$\tau_i$ - lifetime of upper ("initial") level;
$\Delta E$ - energy gap between levels.

Equation (1), being known as two-level model, well describes the experimental lifetime dependence over the whole phase I domain (fig.3, $\tau_{is}(T)$:I) and up to ~300 K for the phase II region.

*Optical properties of the $Bi^+$ center in $KAlCl_4$ - Preprint*     3

In the latter case the significant variation of parameters gives almost the same approximation quality and unambiguous values assignment can't be obtained in this way.

Thereby it was assumed that the leaps in the lifetime are mainly due to changing of energy gap and, as a consequence, the thermalization degree between the levels. Thus it was supposed that the radiative lifetimes of the levels don't change during I↔II phase transition. This assumption doesn't contradict to the experimental results – the two-level model with the same lifetimes but higher energy gap well describes experimental dependence for the phase II up to ~300 K (fig.3, $\tau_{is}(T):II$).

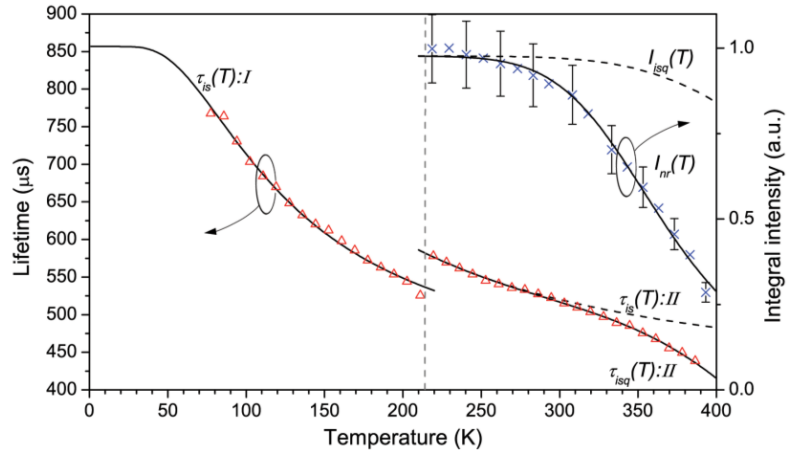

Fig.3. Fitting results of experimental lifetime and integral intensity dependences using equations (1), (2) and (3) respectively. $I_{isq}(T)$-estimation of the luminescence intensity corresponding to the lifetime thermal quenching process. Corresponding functions parameters could be found in table 1. Number of plotted lifetime experimental points was reduced purposely for clarity; approximation was made using all experimental data.

At higher temperatures the luminescence intensity and lifetime decrease, probably, as the result of thermal quenching. Within the frame of single configurational-coordinate model (SCC) while taking into account the Mott model of nonradiative transitions [34] it is possible to write the following equation [33]:

$$\tau_{isq}(T) = \frac{1+e^{-\Delta E/kT}}{\tau_s^{-1} + \tau_i^{-1} \cdot e^{-\Delta E/kT} + \tau_q^{-1} \cdot e^{-(\Delta E+\Delta E_q)/kT}} \quad (2),$$

$\tau_q^{-1}$ - the thermal quenching rate;
$\Delta E_q$ - energy difference between the crossing of adiabatic potential energy curves and the minimum of the excited-state curve.

This model fits well the experimental lifetime data at temperatures above 300 K (fig.3, $\tau_{isq}(T): II$). Using equations (1) and (2) the decreasing of luminescence intensity versus temperature corresponding to the lifetime thermal quenching process could be estimated by the following equation:

$$I(T)=I_0 \cdot \tau_{is}/\tau_{isq} \quad (3)$$

The equation (3) predicts the approximately 15% reduction of luminescence intensity at 400 K while using the parameters from the lifetime fit (fig.3, $Iisq(T)$), but in experiment we observe the ~70% reducing. Nevertheless equation (3) could fit experimental data quite closely with the same $\Delta E_q$ value but much higher quenching rate ($\tau_q^{-1}$) value is needed in this case, indicating that there should exist some additional non-radiative relaxation pathways activated with heating (fig.3, $I_{nr}(T)$).

The corresponding fitting parameters are presented in table 1.

Table 1. Fitting parameters for the lifetime and integral intensity dependences over phases I and II of the crystal for different models represented by Eqs. (1)-(3)

|  | $\tau_s$, μs | $\tau_i$, μs | $\Delta E$, cm$^{-1}$ | $\tau_q^{-1}$, μs$^{-1}$ | $\Delta E_q$, cm$^{-1}$ |
|---|---|---|---|---|---|
| $\tau_{is}(T)$:I |  |  | 167.6 | — | — |
| $\tau_{is}(T)$:II | 856.7 | 249.4 |  |  |  |
| $\tau_{isq}(T)$:II |  |  | 212.1 | 11 | 2572 |
| $I_{isq}(T)$ |  |  |  |  |  |
| $I_{nr}(T)$ |  |  |  | 162 |  |

### 3.3 Emission band analysis

As it was demonstrated above the luminescence spectrum is a superposition of the emission from two different but thermalized levels. Thereby the luminescence band was considered as a sum of two Gaussian components. Not measured luminescence peak shape but its form-function was used in the following analysis because exactly the latter correlates with the emission probabilities [35]. Nevertheless the main obtained conclusions can be extended to the experimentally observed emission band also.

Over the whole temperature range the luminescence band form-function is approximated quite well by a sum of two Gaussians (fig.4); the approximation quality increases with heating particularly for II and III crystal phases.

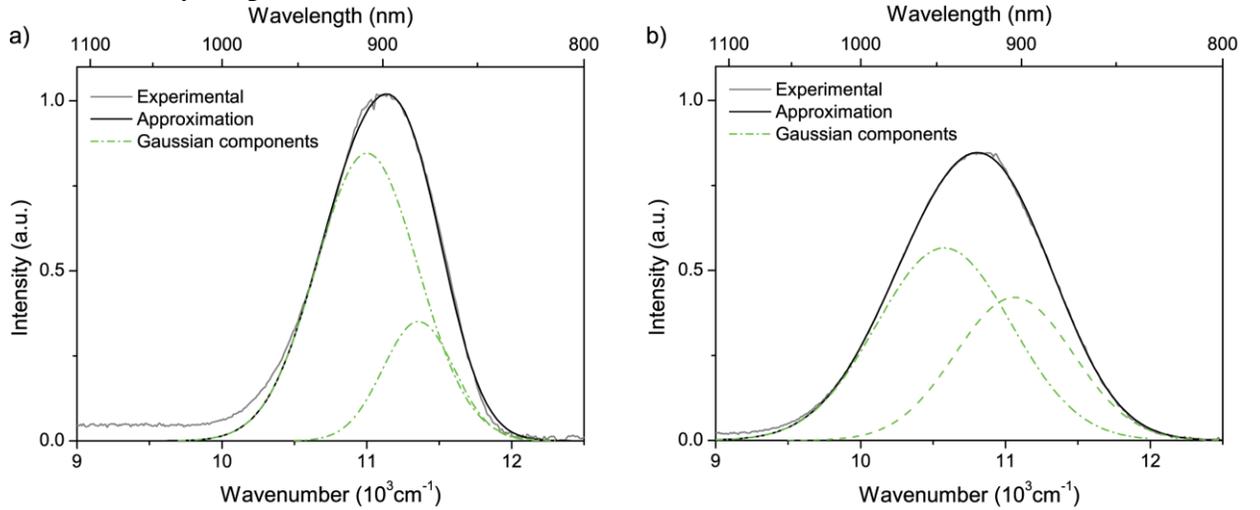

Fig.4. Fitting result of the emission band form-function using double Gaussian approximation at 77 (a) and 202 K (b).

The temperature dependences of the luminescence spectrum parameters: position, FWHM, relative and absolute integral intensity of each Gaussian band were obtained (fig.5).

The relative contribution (fig.5c) of the upper level to the luminescence integral intensity increases with heating in every single crystal phase existence range. The temperature dependences of absolute luminescence intensities starting from different levels (fig.5d) show that contribution of the lower one decreases with heating permanently at the crystal phase II whereas the contribution of the upper one has the maximum at about 300 K and decreases thereafter only, when the thermal quenching is began. Consequently the approximation takes into account the thermalization between the luminescent levels. Moreover this analysis allows to reveal abrupt changes in Gaussian bands parameters during the II↔III phase transition. These facts confirm a physical validity of the used approach.

The theory of the electron-phonon interaction gives the following equations for the first and second moments of the luminescence form function [36, 37]:

$$M_1 = E_m + C \cdot \coth(\hbar\omega/2kT) \qquad (4),$$

$$M_2 = S \cdot (\hbar\omega)^2 \coth(\hbar\omega/2kT) \qquad (5),$$

$S$ – Huang Rhys factor;
$k$ – Boltzmann constant;
$\hbar\omega$ – effective phonon energy.

In the case of Gaussian shape the relation between the second moment and FWHM is expressed by the following equation:

$$M_2 = \frac{FWHM^2}{8\ln 2} \qquad (6)$$

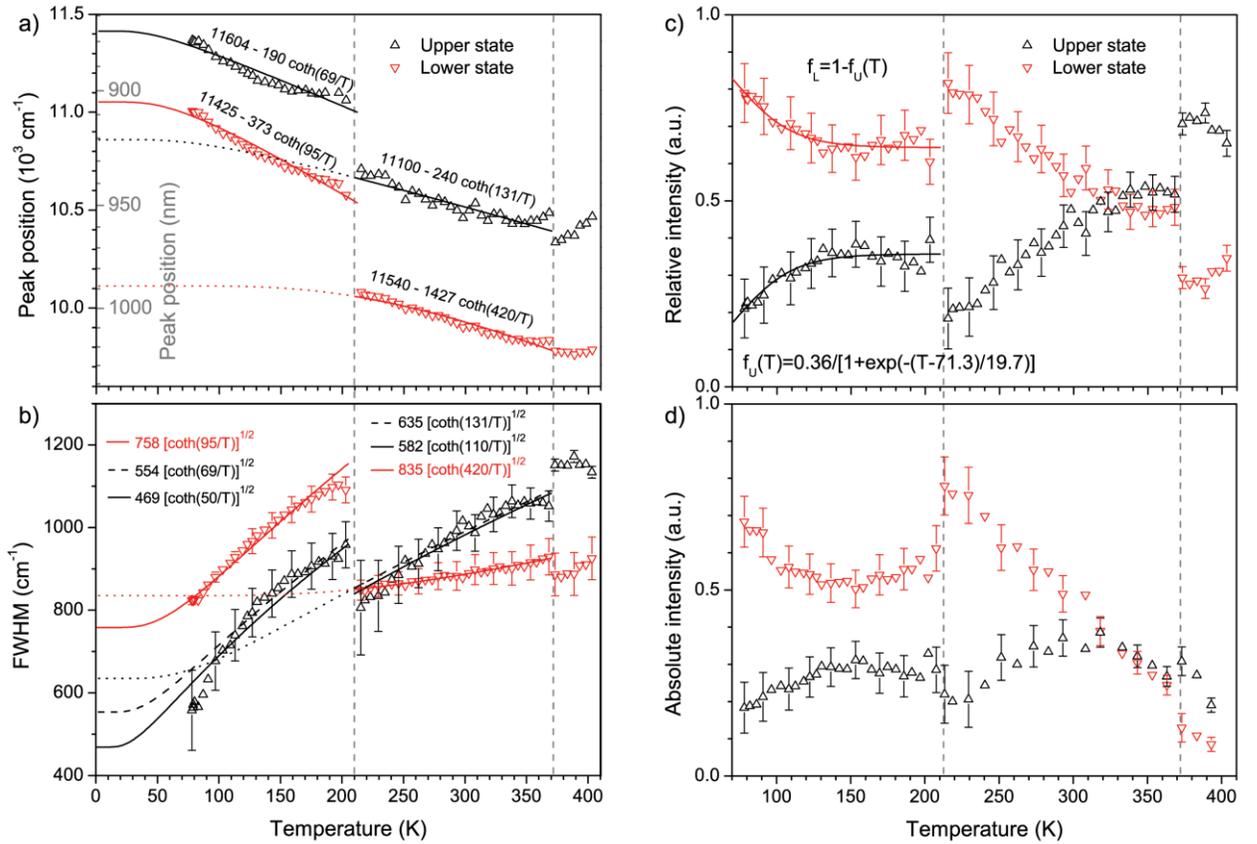

Fig.5. Temperature dependences of peak positions (a), FWHM (b), relative (c) and absolute luminescence intensities (d) corresponding to the components of the luminescence band. Errors in determination of the half-width are defined as inversely proportional to the relative intensity of the band. The positions and FWHM values are extrapolated to 0 K for phase I and phase II (if it would exist at lower temperatures) of the crystal. The latter is shown by doted lines.

Thus the dependences of FWHM over the temperature were approximated by $A(\coth(B/T))^{1/2}$ function for both luminescence levels in existence domains of I and II crystal phases (fig 5.b). Corresponding values of $S$ and $\hbar\omega$ parameters are given in Table 2. Using the all obtained data the set of unambiguous values of $A$ and $B$ parameters can be obtained for the lower luminescence level only. In the case of the upper one the dependence can be approximated with almost the same accuracy by the function with *B<50/110* for I/II phase respectively and only rough estimation of $\hbar\omega$ is possible in this case.

It is important to note that the contribution of upper level integral to the luminescence band is small at low temperatures and doesn't exceed 0.5 up to ~330 K (fig.5c). So it is the most probable reason of the fit ambiguity. If we take into account only the data for which the relative intensity of the upper level contribution exceeds 0.3 (T>100 K and T>250 K temperature regions for I and II crystal phases respectively) then the approximation gives adequately unambiguous parameters values which are close to the upper limit obtained for the all data fitting. Thus the latter values of $\hbar\omega$ and *S* were used for further analysis.

The obtained effective phonon energies were used for fitting of $M_1$ temperature dependence (fig.5a). The corresponding parameters are presented in Table 2.

Table 2. Huang-Rhys parameter and effective phonon energy obtained from the second moment ($M_2$, eq.(5)) temperature dependence fitting and the corresponding C and $E_m$ parameters from the first moment ($M_1$, eq.(4)) fitting for the observed optical bands of $Bi^+$ center in I and II phases of the $KAlCl_4$ crystal.

| Band | Parameter | phase I | | | | phase II | | | |
|---|---|---|---|---|---|---|---|---|---|
| | | $M_2$ | | $M_1$ | | $M_2$ | | $M_1$ | |
| | | S | $\hbar\omega$, cm$^{-1}$ | C, cm$^{-1}$ | $E_m$, $10^3$cm$^{-1}$ | S | $\hbar\omega$, cm$^{-1}$ | C, cm$^{-1}$ | $E_m$, $10^3$cm$^{-1}$ |
| Lum. | *L.*[1] | 5.94 | 132 | -373 | 11.43 | 0.37 | 583 | -1427 | 11.54 |
| | *U.*[1] | 6.02 | 96 | -190 | 11.60 | 2.2 | 182 | -240 | 11.00 |
| Absorption | *Ex.III.b* | 5.63 | 90 | -178 | 12.91 | — | — | — | — |
| | *Ex.III.a* | 1.02 | 168 | 146 | 13.39 | — | — | — | — |
| | *Ex.II.b* | 0.81 | 132 | -44 | 15.45 | 1.35 | 278 | -45 | 15.08 |
| | *Ex.II.a* | 1.54[2] | —[2] | -51 | 15.83 | 0.34 | 459 | 67 | 15.47 |
| | *Ex.I.b* | 1.89 | 126 | 75 | 16.29 | 0.3 | 431 | 232 | 16.34 |
| | *Ex.I.a* | 1.94 | 284 | 859 | 15.65 | 1.04 | 500 | 673 | 16.31 |

[1]The designation "L."("U.") is equal to "band corresponding to the lower (upper) luminescence level"
[2]Due to anomalous temperature behavior of FWHM for *Ex.II.a* band it's impossible to determine effective phonon energy in this case; $\hbar\omega$ is assumed to be equal to the corresponding value for *Ex.II.b* band (132 cm$^{-1}$).

## 3.4. Excitation spectra

The luminescence excitation spectrum over the whole temperature range was measured in the spectral range of 520-700 nm. The data for wavelengths, longer then 700 nm were obtained for I phase only due to low intensity of the luminescence.

As it was shown before there are five well resolved maxima in the luminescence excitation spectrum at 77 K: at 610 (*Ex.I*), 635 (*Ex.II.a*), 650(*Ex.II.b*), 740(*Ex.III.a*) and 790 nm (*Ex.III.b*). All these peaks remains resolved over the whole temperature range corresponding to the crystal phase I. The additional structure of the Ex.II band for II and III crystal phases could be revealed only indirectly by the means of band shape analysis.

The decomposition of the spectra in the wavelength range <700 nm as a sum of two Gaussians (for Ex.II.a and Ex.II.b bands) and one asymmetrical shape (Ex.I) could give precise approximation of experimental data. However the skewness of Ex.I band increases with heating hence it should be considered as sum of several independent bands. Thus the excitation spectra in this range were considered as a sum of four Gaussians. Bands *Ex.III a,b* could be fitted by Gaussians quite well also (fig.6).

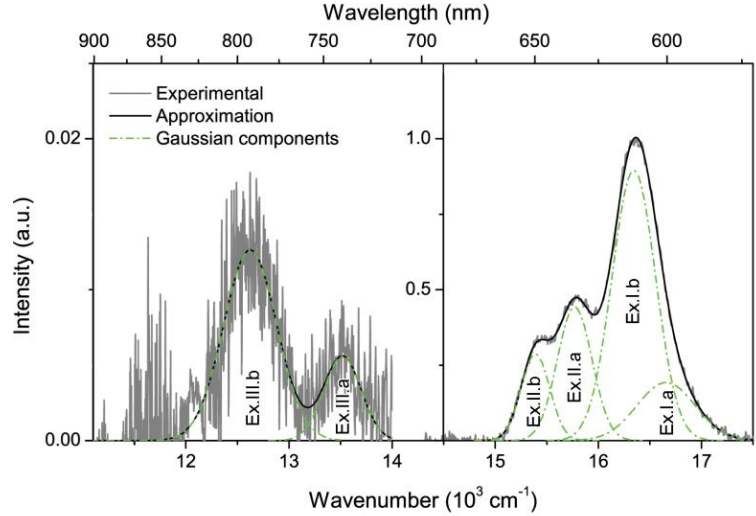

Fig.6. Fitting result of the excitation spectrum at 77 K using six independent Gaussian components. Separate intensity scales corresponding to wavenumbers more and less than 14500 cm$^{-1}$ are depicted for clarity.

Ex.II.a and Ex.II.b bands can not be observed explicitly for the crystal phases II and III [present work, 27]. Thus the most reliable data of the band shape analysis was obtained for the phase I of the crystal. Nevertheless in all the cases used consideration gives the results which at least qualitatively conform to the expected physical pattern during heating.

The analysis of the experimental data gives the main parameters, namely: position, FWHM and relative integral intensity of each Gaussian band versus temperature (fig.7). In the case of excitation spectra the abrupt changes of Gaussian bands parameters during the I↔II phase transition are observed also. Intensity relation of the all excitation bands changes with temperature, especially in the crystal phase I. Relative integral intensity of *Ex.III a,b* bands is about ~0.01 at 77 K. All the bands except *Ex.II* group slightly shift with heating. The positions of the latter ones are almost constant within a single phase domain of the crystal.

Equations (4) and (5) were used for analysis of the excitation spectrum (fitting of FWHM and peak positions) also.

For all the bands the temperature dependences of FWHM were successfully approximated by $A(\coth(B/T))^{1/2}$ equation at crystal phases I and II except Ex.II.a band at phase I. In the latter case the somewhat anomalous behavior is observed. The corresponding values of $S$ and $\hbar\omega$ parameters are given in Table 2.

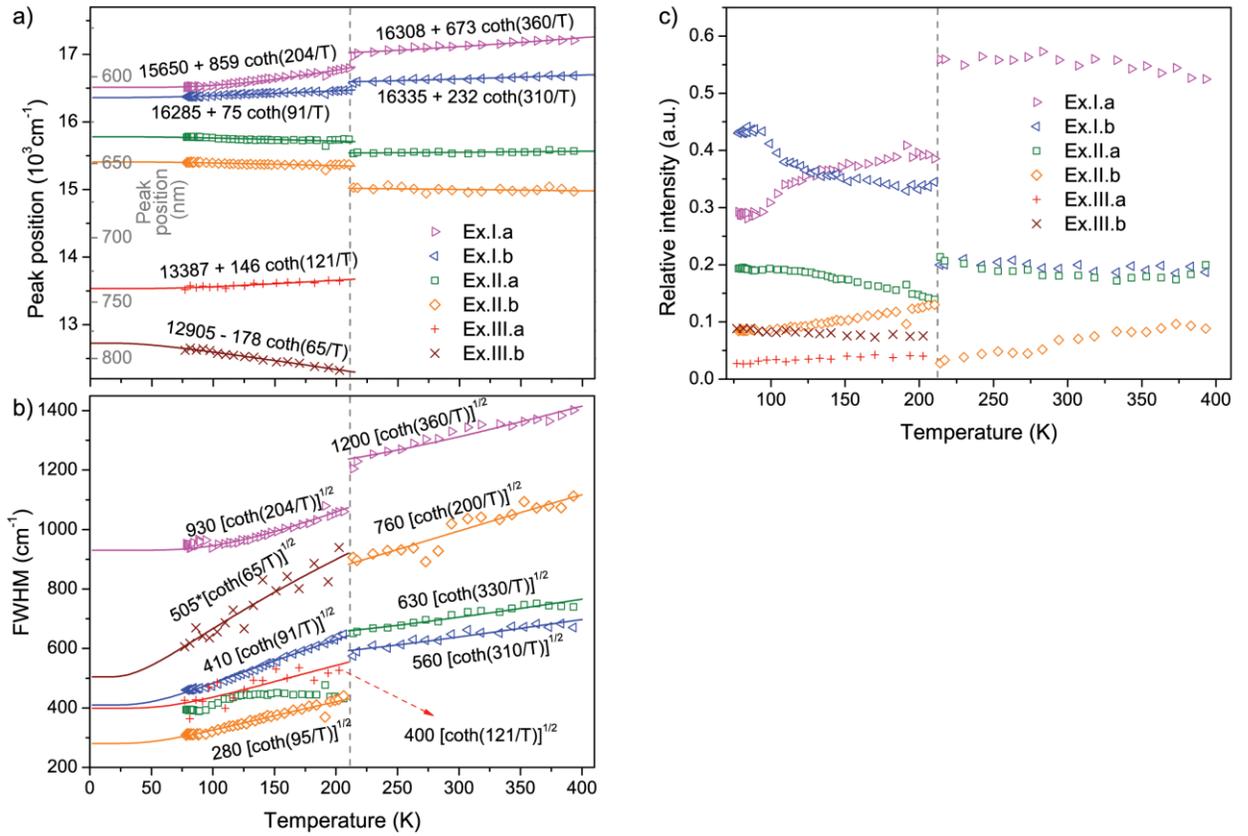

Fig.7. Temperature dependences of peak positions (a), FWHM (b) and relative intensities (c) corresponding to the components of the excitation band. The positions and FWHM values are extrapolated to 0 K for the crystal phase I. Relative intensities of *Ex.III* bands are calculated in assumption of constant contribution of these bands to the integral spectrum being equal to approximately 0.01; these data are enlarged tenfold for clarity.

## *3.5. Analysis of the energy states*

Absorption spectra and energy levels of $Bi^+$ ion in the conformable eutectic molten salt media ($NaCl-AlCl_3$) have been considered in details previously [38, 39]. Besides experimental results the theoretical model in term of ligand field theory, which describes well the observed transitions, was proposed by authors [39].

According to these data we propose to assign the observed optical transitions of $Bi^+$ in $KAlCl_4$ to $^3P_0 \rightarrow {}^3P_2$ ($J_z = \pm 2; \pm 1; 0$) and $^3P_0 \rightarrow {}^3P_1$ ($J_z = \pm 1; 0$) for the absorption bands (see Table 3). Although $J_z$ is not a good quantum number for the eigenstates of $Bi^+$ cation in crystal field surrounding, we still use the above designation of electronic states. However, this designation is rather formal here, because in crystal field the real $Bi^+$ eigenstates of $^3P_1$ or $^3P_2$ manifolds are the mixtures of states with definite $J_z$.

According to electron-phonon theory the temperature dependence of the first moments should be equal for absorption and emission bands of a single same energy state[37]. It is clear that the observed luminescence corresponds to $^3P_1 \rightarrow {}^3P_0$ transition. Comparison of $M_1$ for emission and absorption bands reveals that *Ex.III.b* band corresponds to the emission one. Thus the observed luminescence should be assigned to $^3P_1 (J_z = +1) \rightarrow {}^3P_0$ and $^3P_1 (J_z = -1) \rightarrow {}^3P_0$ transitions for the upper and lower luminescence levels respectively. According to the estimation [39] the energy gap between $^3P_1$ ($J_z = \pm 1$) levels in NaCl-AlCl$_3$ is about 100-300 cm$^{-1}$ that agrees well with our results.

Table 3. The comparison of the energies of Bi$^+$ optical transitions in NaCl-AlCl$_3$ melt (estimation, [39]) and KCl-AlCl$_3$ crystal (experiment, present work). Extrapolated to 0 K values of the absorption transitions energies obtained in the case of the crystal phase I are used here.

| Optical transition | | NaCl-AlCl$_3$ [28] | KCl-AlCl$_3$ | |
|---|---|---|---|---|
| | | Energy, $10^3$ cm$^{-1}$ | Energy, $10^3$ cm$^{-1}$ | Band |
| $^3P_0 \to {}^3P_1$: $J_z =$ | -1 | 11-11.1 | 12.7 | *Ex.III.b* |
| | +1 | 11.2-11.3 | | |
| | 0 | 14.3-14.4 | 13.5 | *Ex.III.a* |
| $^3P_0 \to {}^3P_2$: $J_z =$ | -2 | 15.3 | 15.4 | *Ex.II.b* |
| | +2 | 15.4-15.6 | 15.8 | *Ex.II.a* |
| | 0 | 17 | 16.4 | *Ex.I.b* |
| | ±1 | 17.1-17.4 | 16.5 | *Ex.I.a* |

Unfortunately it is impossible to analyze *Ex.III.b* band shape more accurate to reveal contribution of two peaks corresponding to the luminescence levels due to its low intensity. However it is important to note that temperature behavior of peak positions of the absorption band and the upper luminescence level emission band is almost identical. According to the luminescence lifetime analysis (see above) the transition from the upper luminescent level to the ground state is more allowed than from the lower one. Thus $^3P_0 \to {}^3P_1$ ($J_z = +1$) should contribute more to the *Ex.III.b* absorption band which is in agreement with experiment.

### 3.6. The configuration coordinate model

Consideration of the energy states in terms of the single configuration-coordinate (SCC) model was performed using Franck-Condon approximation and the simplest, quasi-classic, conception. In this model the parameters of the parabolic curves are defined by the following equations:

$$E_a = E_0 + \frac{k_e X_0^2}{2} - \frac{1}{2}\hbar\omega_g \quad (7);\qquad H_e = \left[4\ln 2 \frac{\hbar\omega_e}{k_e}\right]^{1/2} k_g X_0 \quad (9);$$

$$E_e = E_0 - \frac{k_g X_0^2}{2} + \frac{1}{2}\hbar\omega_e \quad (8);\qquad H_a = \left[4\ln 2 \frac{\hbar\omega_g}{k_g}\right]^{1/2} k_e X_0 \quad (10);$$

$H_{a/e}$ – half-width of the absorption/emission band
$\hbar\omega_{g/e}$ – effective phonon energy of the ground/excited state
$E_0/X_0$ – displacement of the excited state by energy/coordinate
$E_{a/e}$ – energy of the absorption/emission transition
$k_{g/e}$ – force constants for the ground/excited state

In the most cases the "one-side" data, i.e. relating to absorption or emission only, were obtained. The found values of Huang-Rhys factor and effective phonon energy allow estimating the Stokes shift by the following equations [40]:

$$\Delta E = \begin{cases} (2S-1) \cdot \hbar\omega, \text{ for } S \geq 1 \\ 2S \cdot \hbar\omega, \text{ for } S < 1 \end{cases} \quad (11)$$

Extrapolated values of the $H$ and $E$ parameters at 0 K are used for the modeling, $\hbar\omega_e$ is considered to be equal to the effective phonon obtained in experiment. For purely quadratic potential curves of the SCC model the equation $\hbar\omega_e / \hbar\omega_g = (H_a/H_{em})^{2/5}$ should be hold [41]. According to the comparison of FHWM for the upper luminescent and *Ex.III.b* absorption bands the values of the effective phonons for $^3P_1$ (Jz = +1) and $^3P_0$ states should be almost equal, i.e. $\hbar\omega_g \approx \hbar\omega_e$ in this case. Thus $\hbar\omega_g = 96$ cm$^{-1}$ is used in the model. Value of $\hbar\omega_g$ can't be obtained by the same way for the phase II and only rough estimation of the SCC potential curves may be done in this case. The ratio of the force constants essentially affects the SCC potential curves while the specific value of $k_g$ depends on the mass of the vibrating system and defines $X_0$ scale only. Some arbitrary units were used for $X_0$ expression in the model.

Table 4. Parameters of the single configuration coordinate curves for the Bi$^+$ center in KAlCl$_4$. In the case of crystal phase I the values for all the energy states were obtained; for the phase II the parameters were estimated for $^3P_1(J_z = \pm 1)$ levels only – the appropriate values are given in brackets

| Energy state | | $k_e/k_g$ | $X_0$, a.u. | $E_0$, 10$^3$ cm$^{-1}$ |
|---|---|---|---|---|
| $^3P_1$: $J_z =$ | -1 | 0.99 (1.51) | 39.45 (25.48) | 11.76 (10.14) |
| | +1 | 0.98 (0.96) | 33.27 (27.77) | 12.05 (11.15) |
| | 0 | 1.81 | 13.58 | 13.33 |
| $^3P_2$: $J_z =$ | -2 | 1.03 | 16.58 | 15.22 |
| | +2 | 1.43 | 16.75 | 15.53 |
| | 0 | 1.34 | 18.81 | 16.08 |
| | ±1 | 2.4 | 23.76 | 15.78 |

The calculated SCC diagrams are presented on fig.8, the corresponding parameters of these models could be found in Table 4. The curves for $^3P_2$ levels for the crystal phase II are strongly depend on $\hbar\omega_g$ value thereby in this case the estimation was made for $^3P_1$ excited states only.

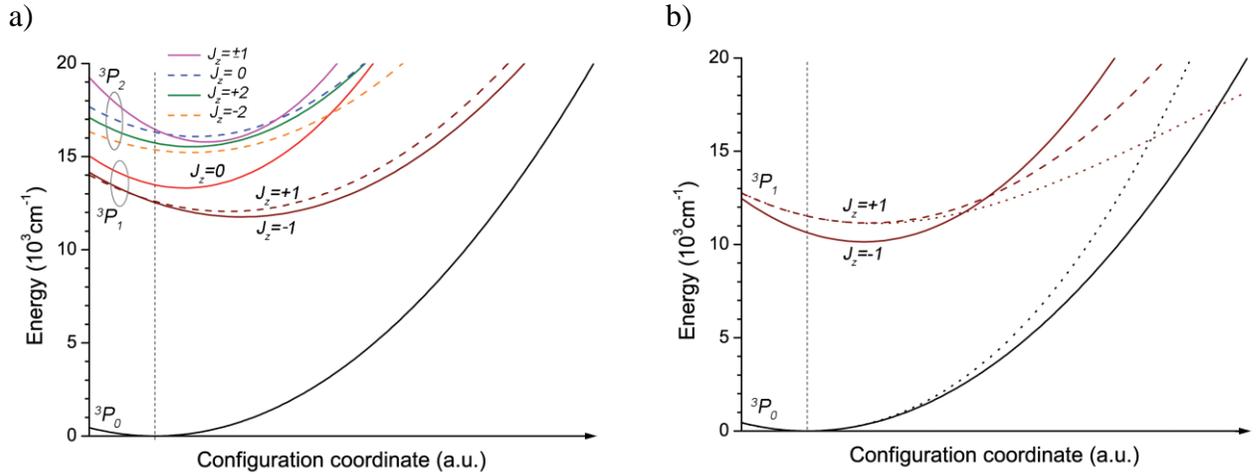

Fig.8. The single configuration coordinate model of $Bi^+$ optical center in $KAlCl_4$ phases I (a) and II (b). Only the curves for $^3P_1$ excited states were obtained for the crystal phase II due to lack of experimental data. The dotted lines on the crystal phase II SCC diagram refer to probable non-harmonic SCC curves.

## *4. Discussion*

Further investigation of the $Bi^+$ optical centre in $KAlCl_4$ demonstrates that abrupt changes of the main luminescent properties occur during two phase transitions of the crystal. In the case of the luminescence lifetime both of these changes could be observed directly in experiment whereas the additional analysis is needed to detect the luminescence band parameters leaps in the case of II↔III phase transition.

The existence of the phase transitions in $KAlCl_4$ host requires to consider each state of the crystal independently. This condition narrowed significantly the temperature range available for the analysis, complicated and decreased the accuracy of the consideration. Nevertheless quite consistent and certain results are obtained. The presence of the two luminescent levels revealed in the lifetime and the luminescence intensity analysis. This fact found the additional validation in course of emission band shape consideration: uncommon temperature behavior of the peak position and FHWM could be understood in terms of two thermalized levels and described as the result of changes in thermal equilibrium population of the levels.

Analysis of the temperature dependences demonstrates quantitatively the significant difference in the electron-phonon interaction of $Bi^+$ impurity center with the host between its different polymorphic phases – the values of effective phonon energy and Huang-Rhys factor vary considerably in different crystal states. The changes in the local surrounding of the center are the most probable reason for these observations.

The observed changes can be clearly seen when comparing the SCC diagrams for I and II crystal phases: the position of the curve minimum as well as the curvature relationships of SCC curves for the different polymorphic phases are quite different. It should be noted that SCC diagrams predict the increased energy gap between $^3P_1 (J_z = \pm 1)$ levels for the crystal phase II (as compared to phase I), and thus partly confirms the above used assumption that have been made during the analysis of $Bi^+$ luminescence lifetime in the case of $KAlCl_4$ phase II.

In terms of considered conception the energy curves present parabolas which is the significant simplification of the physical reality. For instance the harmonic approximation cannot reproduce the temperature shift of peak maxima [42]. Furthermore the phase II SCC diagram doesn't predict significant thermal luminescence quenching. Taking into account the anharmonicity one could probably reveal the missing details of the picture. For example, the addition of some cubic terms could decrease energy difference between the excited state minimum and its crossing point with the ground state, making the crossing point easily attainable by thermal fluctuations (fig.8.b). Nevertheless the SCC diagram for the crystal phase I shows that sublevels of $^3P_2$ term are very close to each other allowing non-radiative relaxation to the lowest of these states. Also, the non-radiative relaxation pathway from $^3P_2$ to $^3P_1 (J_z =\pm 1)$ sublevels via intermediate $^3P_1 (J_z =0)$ is highly probable. The absence of this channel probably would result in additional luminescence band(s) in red region (~620-670 nm) however we didn't observe it experimentally in this crystal.

The calculated Huang-Rhys parameter values of the luminescent levels at the crystal phase I indicate that there is a strong electron-phonon coupling here and even at lower temperatures zero-phonon line could be hardly observed. Obtained values are similar e.g. with ones for $Tl^0$ impurity and less, than typical values for F-centers in alkali halides [43 and References inside, 40]. This is the one more evidence that exactly impurity center (namely $Bi^+$ ion), but not a defect, is responsible for the observed emission.

Even lower values of $S$ are obtained for energy states, which are responsible for the $Bi^+$ absorption in $KAlCl_4$ phases I, II, and the luminescent levels in the crystal phase II. Thus it could be possible to observe zero-phonon line for $^3P_0 \rightarrow {}^3P_2$ and $^3P_0 \rightarrow {}^3P_1 (Jz=0)$ optical transitions.

It could be seen that the optical properties of $Bi^+$ center strongly depend on the host media and varies significantly even in different polymorphic phases of the same crystal. Thus it is difficult to compare results being obtained for different materials. Nevertheless the observed differences in $Bi^+$ absorption and emission characteristics between various media are not as dramatic as one could expect. The comparison of $Bi^+$ optical features in $KAlCl_4$, $KMgCl_3$, $RbPb_2Cl_5$ and the chloride glass shows considerable similarity between these materials [26]. In particular the results obtained for $RbPb_2Cl_5$ should be noted – some narrow lines were observed in absorption spectra at 4.7 K by authors [11]. Probably these lines could correspond to zero-phonon transitions.

We believe that exactly $Bi^+$ optical center is often responsible for the emission near 1μm in other Bi-doped media too. Considerable similarities was found when comparing silica, germanate and chloride glasses [23]. In recent investigation of Ga/Bi co-doped silica glass the model of $Bi^+$ and associated defect was proposed for the explanation of observed photoluminescence [44]. In this paper the estimation of the upper limit of Huang-Rhys parameter for the lowest excited state of $Bi^+$ $S<0.5$ was made by the authors. It is in a good agreement with $S=0.37$ value obtained here for $^3P_1 (Jz=-1)$ state in the case of the crystal phase II.

Similar results concerning the dual structure of the luminescence band were obtained previously by Dvoyrin et al. [45]. The authors observed the blue shift of the luminescence band near 1.1 μm at heating in Bi-doped silica fiber and it was explained by variation of luminescence intensities strating from two different energy terms belonging to the same active center. The difference in emission energy was 570 cm$^{-1}$ in this work. These findings are rather similar to the results obtained in our experiment.

## 5. Summary

The optical properties of the Bi$^+$ substitutional center in KAlCl$_4$ crystal are studied in details. The strong dependence of the Bi$^+$ center features on its local surrounding is demonstrated experimentally. The two independent but closely spaced energy states of the ion are responsible for the observed emission near 1 μm. The observed temperature dependences of the emission lifetime and the band shape are determined mainly by thermalization process, which takes place between these levels. The optical transitions were assigned to the specific energy states of Bi$^+$: $^3P_0 \rightarrow {}^3P_2$ ($J_z = \pm 2; \pm 1; 0$) and $^3P_0 \rightarrow {}^3P_1$ ($J_z = \pm 1; 0$) for the absorption and $^3P_1$ ($J_z = \pm 1$) $\rightarrow {}^3P_0$ for the emission. The Huang-Rhys parameter and the effective phonon energy values are determined for these states by investigation of temperature behavior of the observed transitions. The obtained values of Huang-Rhys parameter indicate on the impurity, not defect, nature of the optical center. The configurational coordinate model of Bi$^+$ impurity in KAlCl$_4$ crystal is proposed. Probably, the non-radiative relaxation between optically excited $^3P_2$ and luminescent $^3P_1$ ($J_z = \pm 1$) states proceed via $^3P_1$ ($J_z = 0$) energy level which lies between them. Brief comparison of the obtained results with other Bi-doped materials shows that despite strong dependence of the Bi$^+$ optical properties on the surrounding some similarities could be found quite often. Thus the results and conclusions being obtained for Bi$^+$ in KAlCl$_4$ crystal could be used with appropriate corrections for other Bi-doped media, in particular for the identification of the observed NIR emission.

*Acknowledgments*

The reported work was supported by the Russian Foundation for Basic Research (RFBR, research project number 13-03-00777)